\documentclass[%
 reprint,
 amsmath,amssymb,
 aps,
 prl,
]{revtex4-2}

\usepackage{graphicx}
\usepackage{dcolumn}
\usepackage{bm}
\usepackage{color}

\usepackage{cleveref}
\usepackage{afterpage}
\usepackage{float}
\usepackage{amsmath}
\usepackage{bm}

\begin{document}

\preprint{APS/123-QED}

\title{Hydrodynamic synchronisation of chiral microswimmers}

\author{Sotiris Samatas}
\affiliation{%
 Univ. Bordeaux, CNRS, LOMA, UMR 5798, F-33400 Talence, France 
}

\author{Juho Lintuvuori}%
\affiliation{%
 Univ. Bordeaux, CNRS, LOMA, UMR 5798, F-33400 Talence, France 
}%

\date{\today}

\begin{abstract}

We study synchronization in bulk suspensions of spherical microswimmers with chiral trajectories using
large scale numerics. The model is generic. It corresponds to the lowest order solution of a general model
for self-propulsion at low Reynolds numbers, consisting of a nonaxisymmetric rotating source dipole. We
show that both purely circular and helical swimmers can spontaneously synchronize their rotation. The
synchronized state corresponds to velocity alignment with high orientational order in both the polar and
azimuthal directions. Finally, we consider a racemic mixture of helical swimmers where intraspecies
synchronization is observed while the system remains as a spatially uniform fluid. Our results demonstrate
hydrodynamic synchronization as a natural collective phenomenon for microswimmers with chiral
trajectories.

\end{abstract}

\maketitle

\paragraph*{Introduction.---}

Microswimmers are a subset of active matter systems and correspond to microscopic elements self-propelling within a fluid environment. Natural microswimmers consist of biological microorganisms~\cite{koch2011collective,lauga2016bacterial,lauga2020fluid} and their collective dynamics has gained a lot of interest of late~\cite{dombrowski2004self, sokolov2007concentration, wolgemuth2008collective, cisneros2010fluid, zhang2010collective, sokolov2012physical, wensink2012meso, dunkel2013fluid, peng2021imaging}. This has inspired research on synthetic microswimmers, typically based on phoretic Janus particles~\cite{ebbens2018catalytic,theurkauff2012dynamic,ginot2018aggregation}. The interest for developing artificial swimmers has been fuelled by the various promissing possibilities for applications such as micro-cargo transportation\cite{baraban2012transport, boymelgreen2018active, mena2021polymeric, bunea2020recent}, targeted drug delivery\cite{srivastava2016medibots, singh2019multifunctional, bhuyan2017magnetic, bunea2020recent}, artificial insemination\cite{medina2016cellular, bunea2020recent} and microsurgery\cite{nelson2010microrobots, srivastava2016medibots, bunea2019strategies, york2021microrobotic, bunea2020recent}.

Most theoretical studies of microswimmer suspensions have concentrated on particles that swim in straight lines, with simulations predicting the spontaneous formation of collective swimming along a common direction --- uniform polar order~\cite{evans2011orientational, alarcon2013spontaneous, yoshinaga2017hydrodynamic, yoshinaga2018hydrodynamic, delmotte2015large, theers2018clustering, oyama2017hydrodynamically}. However, microorganisms typically have intrinsic chirality and tend to swim along helical paths~\cite{bray2000cell, jennings1901significance, jekely2008mechanism, fenchel1999motile, mchenry2003kinematics, marumo2021three, corkidi2008tracking, jikeli2015sperm, thar2001true, su2013sperm,brumley2012hydrodynamic}. Similarly, any asymmetry due to imperfections in the shape of the colloids or in their catalytic coating would also lead to chiral motion for artifical swimmers~\cite{vilela2017microbots, lancia2019reorientation, brown2014ionic, ebbens2018catalytic, campbell2017helical}. 

Continuum descriptions based on the long-range hydrodynamics produced by flow singularities~\cite{blake1974fundamental, pozrikidis1992boundary, pak2015theoretical, lauga2009hydrodynamics} have been extensively used in the past, with some works including chiral flows~\cite{furthauer2012active, friedrich2009steering, singh2018generalized}. However, these models fail to capture near-field hydrodynamic effects, which are believed to be crucial for the formation of polar order~\cite{yoshinaga2017hydrodynamic,yoshinaga2018hydrodynamic}. 

Most of the current theoretical work of active particles moving along chiral paths relies on dry microscopic descriptions such as active Brownian particle (ABP) models~\cite{lowen2016chirality, liebchen2017collective, levis2019simultaneous, liao2018clustering, levis2018micro, bickmann2020analytical, lei2019nonequilibrium, ma2022dynamical, liao2021emergent,liebchen2022chiral}. These effectively account for excluded volume effects, but neglect hydrodynamic interactions. Simulations of rotational dry models have predicted large-scale synchronisation, when a Kuramoto-type alignment term is included~\cite{liebchen2017collective,levis2019activity}.
Very recently, work on the hydrodynamics of chiral swimmers has started to emerge, but has so far been limited to single and two particle systems \cite{fadda2020dynamics, burada2022hydrodynamics, maity2022near, maity2021unsteady, rode2021multi,lisicki2018autophoretic}. 

Explicitly incorporating chirality in hydrodynamic models used to study microswimmer suspensions could have an important effect regarding the emergence of collective states, such as large-scale collective oscillations\cite{chen2017weak, zhang2020oscillatory}, polar order~\cite{evans2011orientational, alarcon2013spontaneous, yoshinaga2017hydrodynamic, yoshinaga2018hydrodynamic, delmotte2015large, theers2018clustering, oyama2017hydrodynamically} or hydrodynamic synchronisation~\cite{putz2009hydrodynamic,qian2009minimal,uchida2010synchronization, kotar2010hydrodynamic, golestanian2011hydrodynamic, theers2013synchronization, han2020reconfigurable}. 
While synchronisation arising from active flows has been predicted for linear trimers~\cite{putz2009hydrodynamic} and for rotors on a 2-dimensional lattice~\cite{golestanian2011hydrodynamic}, the ability of microswimmers to spontaneously synchronise (or not) in freely moving bulk suspensions, remains an open question. 

Here, we show that swimmers with chiral trajectories can synchronise their rotation in a fully three-dimensional suspension. We consider finite sized swimmers, with a surface slip-flow  arising from the general solution for self-propulsion at low Reynolds numbers~\cite{pak2014generalized}, corresponding to a rotating source dipole flow inclined at an angle $\psi$ with respect to the particle polar direction. A synchronised state, corresponds to the alignment of these dipoles.
We study three distinct cases: circular swimmers, helical swimmers, and a racemic mixture of left-handed and right-handed helical swimmers. In all cases, the spontaneous formation of synchronised states is observed. 

\begin{figure}[t!] 
\includegraphics[width=0.48\textwidth]{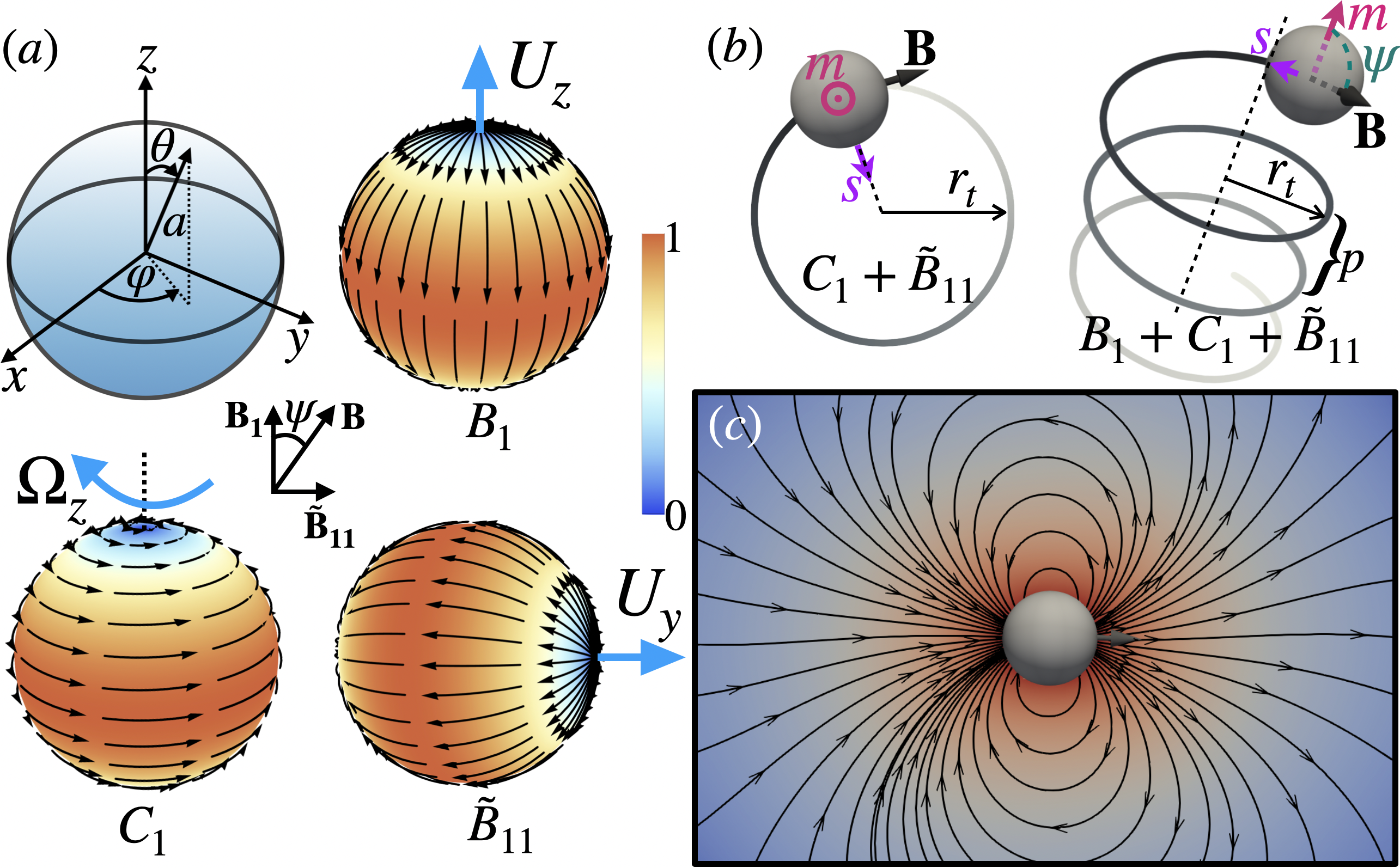}
\caption{Model for rotational squirmers. (a)  The surface slip-flows corresponding to the different modes: $B_1$, $C_1$ and $\widetilde{B}_{11}$ in the particle frame. The magnitude of the {normalised} surface velocity (slip flow) for each mode is represented by a colour-code and the streamlines are coloured black.  (b) The particle trajectories in the lab frame, corresponding to circular (left) and helical (right) swimming. The unit vectors $\bm{m}$ and $\bm{s}$ correspond to the particle polar and azimuthal axes respectively and $\psi$ is the inclination angle with respect to $\bm{m}$. (c) Swimmer flow field obtained from the simulations, corresponding to a source dipole $\mathbf{B}$ (neutral squirmer). The magnitude of the fluid velocity is coloured using a logarithmic scale and overlaid by black streamlines.}
\label{fig1}
\end{figure}

\paragraph*{Model for rotational squirmers.---}

To model the microswimmers, we consider spherical particles of radius $a$,  and extend the standard squirmer model~\cite{lighthill1952squirming,blake1971spherical} to include rotational slip-flows. Based on Lamb's general solution, the tangential slip-flow at the particle surface, is given in spherical coordinates by an infinite series of modes for the polar and azimuthal components $\mathbf{e}_\theta$ and $\mathbf{e}_\varphi$\cite{pak2014generalized}. The lowest order modes correspond to self-propulsion (source dipoles and rotlets), while the higher order terms correspond to fluid mixing. We choose~\cite{pak2014generalized,SupMat}
\noindent
\begin{equation} \label{eq4}
\begin{gathered}
     u_\theta\rvert _{r=a} = B_1\sin\!\theta + \widetilde{B}_{11}\cos\!\theta\sin\!\varphi\\
     u_\varphi\rvert _{r=a} = C_1\sin\!\theta + \widetilde{B}_{11}\cos\!\varphi . \ \ \ \ \ \ 
\end{gathered}
\end{equation}
\noindent
The $B_1$ mode corresponds to the source dipole in the standard squirmer model (top right panel in Fig.~\ref{fig1}a). $C_1$ leads to a rotation of the particle around its polar axis $\bm{z}$ (or $\bm{m}$) with an angular velocity $\omega_0=C_1/a$ (bottom left panel in Fig.~\ref{fig1}a). $\widetilde{B}_{11}$ corresponds to a source dipole along $\bm{y}$ (bottom right panel in Fig.~\ref{fig1}a). 
 The total swimmer flow field corresponds to a single source dipole $\mathbf{B}$ with magnitude $B=\sqrt{\widetilde{B}_{11}^{2} + B_1^2}$, which rotates around the polar axis ($\bm{m}$) at an inclination $\psi = |\tan^{-1}(\widetilde{B}_{11}/B_1)|$ (Fig.~\ref{fig1}).  An isolated particle has a swimming speed $v_0=\frac{2}{3}B$.  When $\psi =90^{\circ}$ ($B_1 = 0$) the swimmers have circular trajectories in a plane perpendicular to $\bm{m}$ (left in Fig.~\ref{fig1}b). The radius of the trajectory is given by $r_t=2\widetilde{B}_{11}a/(3C_1)$ and the period by $T_0=2\pi/\omega_0=2\pi a/C_1$. 
For $\psi \neq 90^{\circ} $ and $\psi \neq 0^{\circ} $ the trajectories become helical with  pitch length $p = 4\pi B_1 a/(3C_1)$ (right panel in Fig.~\ref{fig1}b). To characterise the helical swimming, we define the ratio $\lambda \equiv r_t/p = \widetilde{B}_{11}/(2\pi B_1)$~\cite{SupMat}.

To study the collective dynamics of suspensions of $N$ swimmers, we use the lattice Boltzmann method~\cite{SupMat}.
The typical particle Reynolds number is $\mathrm{Re}\sim 0.01$ with simulation times $\sim 100$s. (see supplementary material~\cite{SupMat} for details of simulations and mapping to SI units). An orientationally ordered state, corresponds to the alignment of the source dipoles $\mathbf{B}$. The amount of alignment can be measured by considering a velocity order parameter $P_{v}(t) = \frac{ \left\lvert \sum_{i}^{N}\bm{\hat{v}}_i \right\rvert}{N}$, where $\bm{\hat{v}}_i= \bm{v}_i/{v}_i$. To further quantify the ordering, we measure the alignment along the azimuthal $\bm{s}$ and polar $\bm{m}$ directions, by calculating
 $P_{{s}|{m}}(t) = \frac{ \left\lvert \sum_{i}^{N}\bm{s}_i|\bm{m}_i \right\rvert}{N}$. $P_{{{v}}|{s}|{m}} = 1$ corresponds to complete  order, and $0$ to an isotropic state.

\begin{figure}[t!] 
\includegraphics[width=0.48\textwidth]{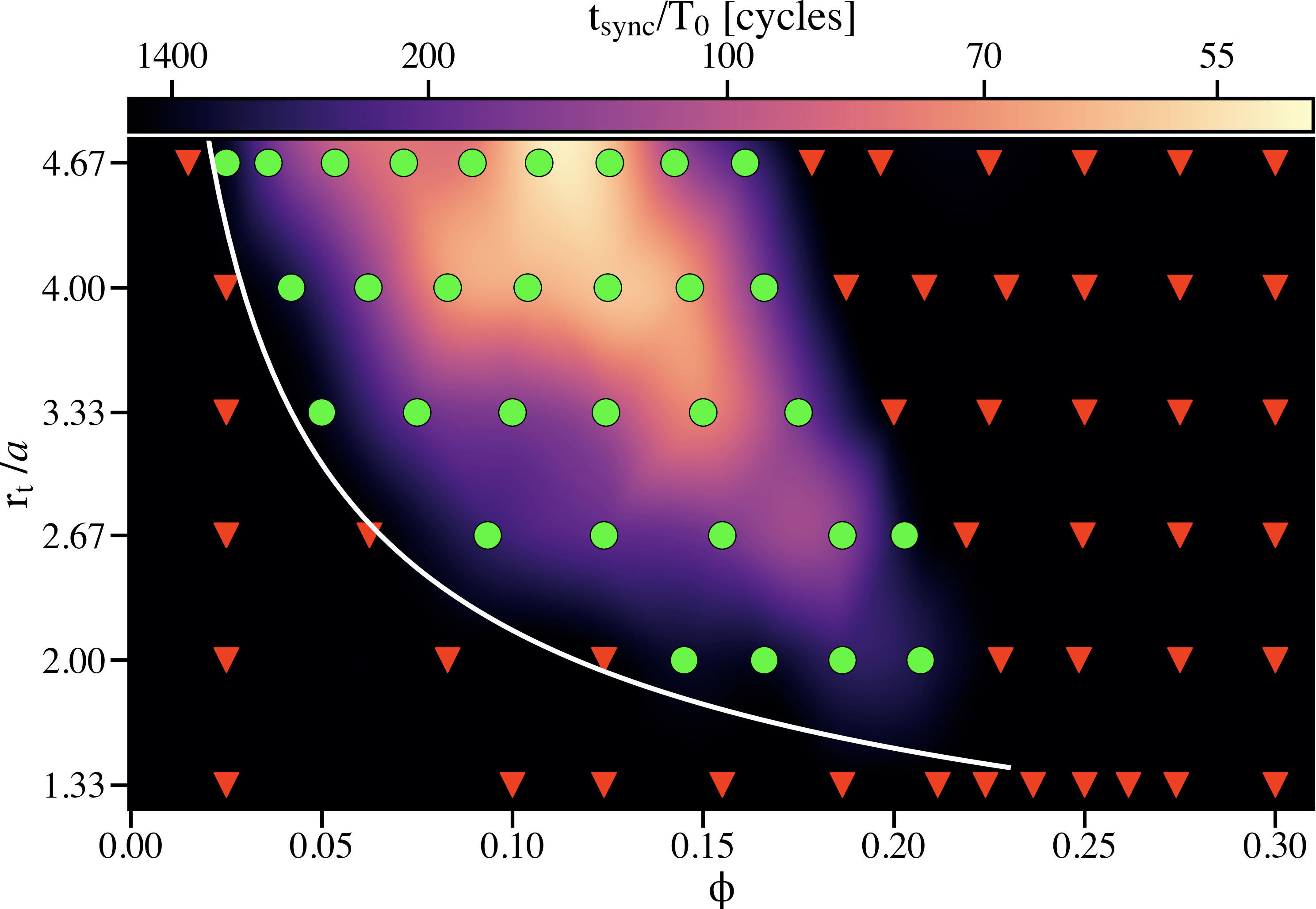}
\caption{Synchronisation diagram for circular swimmers $(\psi = 90^\circ)$ as a function of the volume fraction $\phi$ and particle trajectory radius $r_t$. Green circles indicate global synchronisation, and the red triangles mark isotropic states. The synchronisation region is coloured according to a waiting time $t_{sync}/T_0$ corresponding to the total time elapsed from the start of the simulation until synchronisation is reached. The white curve corresponds to $\phi = \phi'_c\tfrac{4/3\pi a^3}{2\pi r_t^2 a}$ with $\phi'_c = 70$\%. (see text and supplementary material \cite{SupMat} for details).}
\label{fig2}
\end{figure}

\paragraph*{Synchronisation of circular swimmers.---}

\begin{figure}[t!] 
\includegraphics[width=0.48\textwidth]{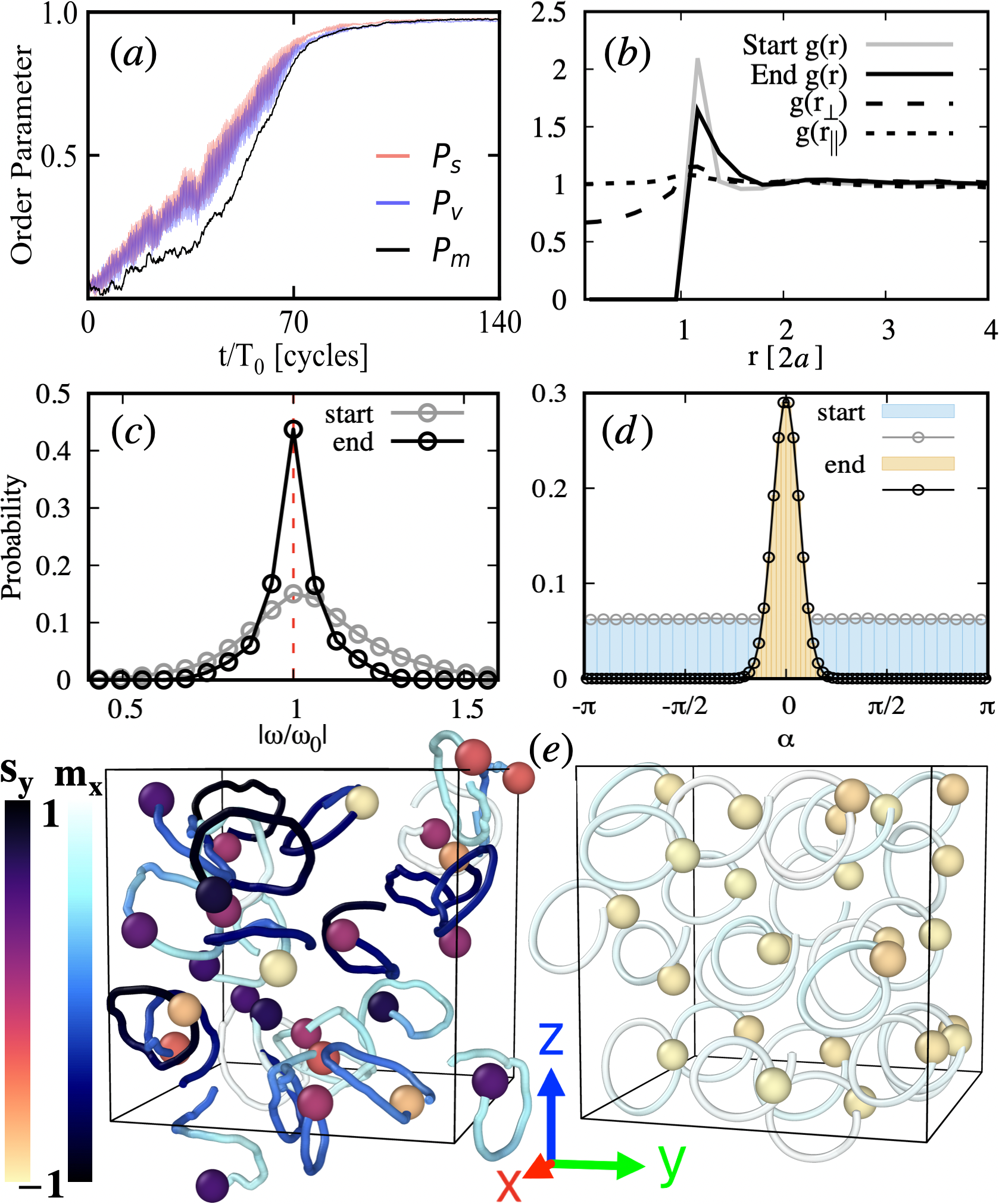}
\caption{ (a) Circular swimmers: Example of a typical time evolution of the azimuthal $P_s$ (red), velocity $P_v$ (blue) and polar $P_m$ (black) order parameters. 
(b) Radial distribution function $g(r)$ of the system at the beginning (gray) and at the end (black) of the simulation. The $g(r_{\perp})$ ($g(r_{||})$) are calculated perpendicular (parallel) to the polar director $\bm{P}_M$.
Probability distribution of the (c) angular velocities $\omega$ and (d) phase lag angle $\alpha$ between all particle pairs, at the start and end of the simulation. 
(e) Snapshots of $25$ selected particles at the beginning (left) and end (right) of the simulation.  The particles are coloured according to the $y-$component of their $\bm{s}$ vector. The trajectories are shown over one period and coloured according to the $x-$component of the swimmer's $\bm{m}$ vector. (The data corresponds to $\phi\approx 0.15$ and $r_t\approx 3.33a$).}
\label{fig3}
\end{figure}

Starting from isotropic initial conditions, we find that circular swimmers spontaneously synchronise their rotation when $\phi\approx 3\ldots 23$\% and $r_t\approx 2\dots 5a$ (Fig.~\ref{fig2}). The synchronisation corresponds to the spontaneous alignment of the particle velocities, with the growth of both azimuthal and polar order, where typically $P_s\approx P_m\approx P_v\gtrsim 0.85$ at long times (Fig.~\ref{fig3}a). 
The phase locking is apparent from the distribution of the lag angle $\alpha = \alpha_{1,2}^{s\perp}$ calculated from all the particle pairs, considering the $\bm{s}$ vectors of two different rotors in the plane perpendicular to the global polar director, $\bm{P}_M\sim\sum_{i}^{N}\bm{m_i}$.  The distribution of $\alpha$ changes from uniform at $t\approx 0$ to a normal distribution with a peak at $\alpha\approx 0$ in the globally synchronised state (Fig.~\ref{fig3}d). 
In this state, the particle trajectories are circular and aligned perpendicularly to $\bm{P}_M$ 
(right in Fig.~\ref{fig3}e). The particle positions remain isotropic  with the pair-correlation functions $g(r)$, $g(r_{\perp})$ and $g(r_{||})$ showing liquid-like structure (Fig.~\ref{fig3}b).

\begin{figure}[t!] 
\includegraphics[width=0.48\textwidth]{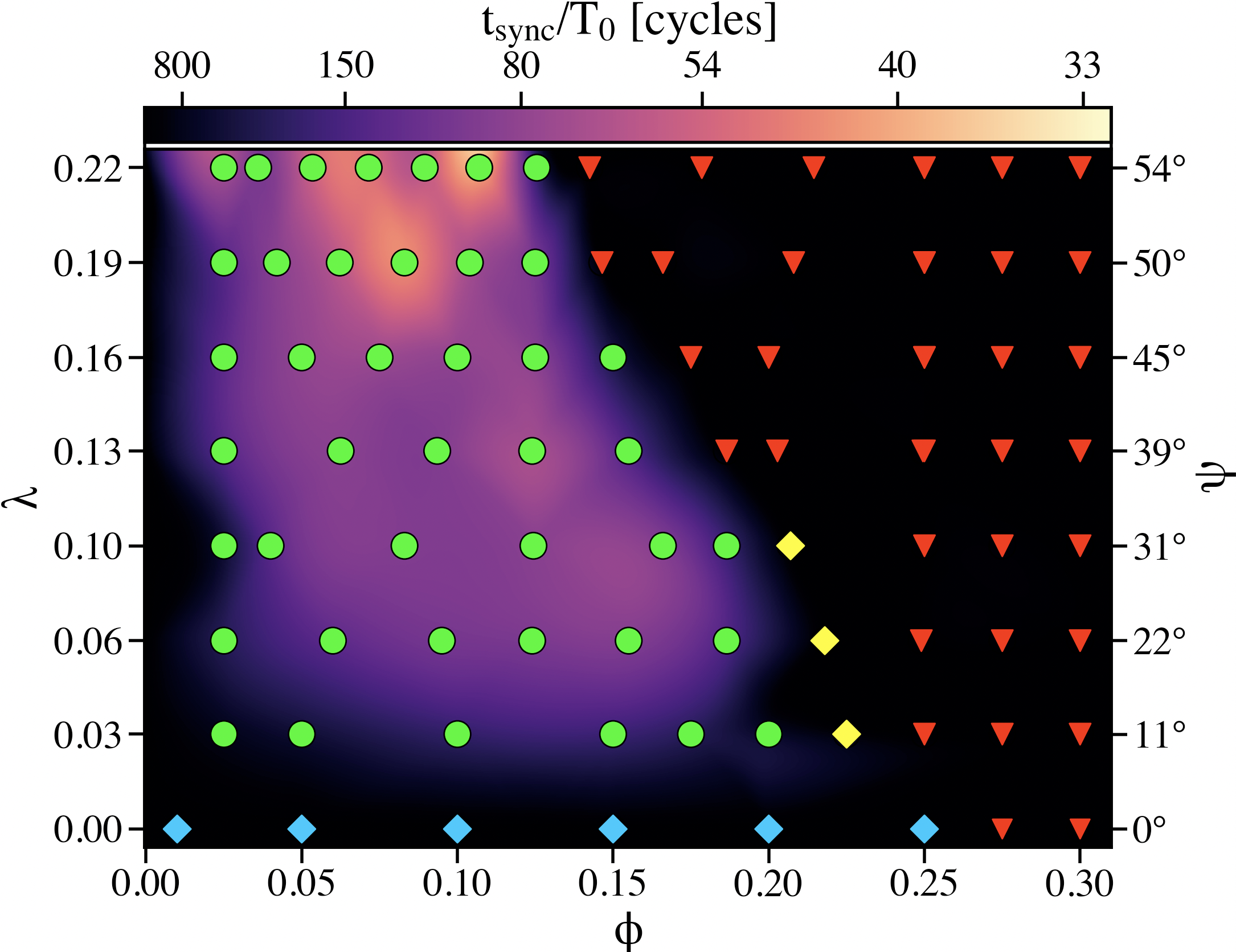}
\caption{State diagram for helical swimmers as a function of $\phi$ and $\lambda = r_t/p$. The green circles correspond to global synchronisation, and red triangles to isotropic states. 
The blue diamonds mark polar order for classic linear neutral squirmers and yellow diamonds correspond to finite polar order in the absence of synchronisation for chiral swimmers. (data corresponds to $p\approx 21 a$).}
\label{fig4}
\end{figure}

The likelihood of the synchronisation depends on the volume fraction $\phi$ and the trajectory radius $r_t$ (Fig.~\ref{fig2}). At low $\phi$ 
the system remains in an isotropic state with the circular trajectories randomly oriented and distributed. 
When $\phi$ is increased, the trajectories become jagged in the isotropic state (left in Fig.~\ref{fig3}e).  At long times the trajectories align (right in Fig.~\ref{fig3}e). The distribution of rotational frequencies $\omega$ has a peak at $\omega_0$ and the width likely arises from the hydrodynamic fluctuations (Fig.~\ref{fig3}c). Interestingly, the particle dynamics is reminiscent of the active-absorbing state transition predicted for dry circular swimmers in 2-dimensions (2D)~\cite{lei2019nonequilibrium} --- the diffusive dynamics in the isotropic state becomes sub-diffusive when the spontaneous synchronisation occurs~\cite{SupMat}. However, in the 2D dry system, where the particles interact exclusively via steric collisions, only local synchronisation was observed~\cite{lei2019nonequilibrium}. This suggests that hydrodynamic interactions are crucial for the large scale synchronisation observed here.
 
 Previous studies of linear squirmers  predict that the alignment of source dipoles corresponding to the formation of uniaxial polar order is dominated by near-field hydrodynamic interactions~\cite{yoshinaga2017hydrodynamic,yoshinaga2018hydrodynamic}. 
When $r_t\sim a$, an isolated swimmer sweeps an area $\sim r_t^2$ during one period $T_0$, and  can be thought to occupy an effective volume $2\pi r_t^2 a$. The lower-$\phi$ limit for the synchronisation region, closely corresponds to the random close packing of discotic cylinders with volume $2\pi r_t^2 a$ ~\cite{SupMat} (white line in Fig.~\ref{fig2}).
Above this line, the effective volumes overlap in the isotropic state, and the swimmers have a high probability of interacting via near-field hydrodynamics. 

To study the ordering dynamics, we measure the total time $t_{sync}$ from the beginning of the simulation until synchronisation is reached . 
The fastest formation is observed in the middle of the synchronised region (Fig.~\ref{fig2}).
For a given $r_t$, if $\phi$ is too large no synchronisation is observed. This implies the existence 
of a dynamic bottleneck where the particles have multiple collisions during their full-rotation time $T_0$, hindering the growth of global alignment. 
For simulations towards the high-$\phi$ end of the synchronisation region, $t_{sync}$ is increased (Fig.~\ref{fig2}), and the order parameters fluctuate close to zero before the growth of the order begins. 

\begin{figure}[t!] 
\includegraphics[width=0.48\textwidth]{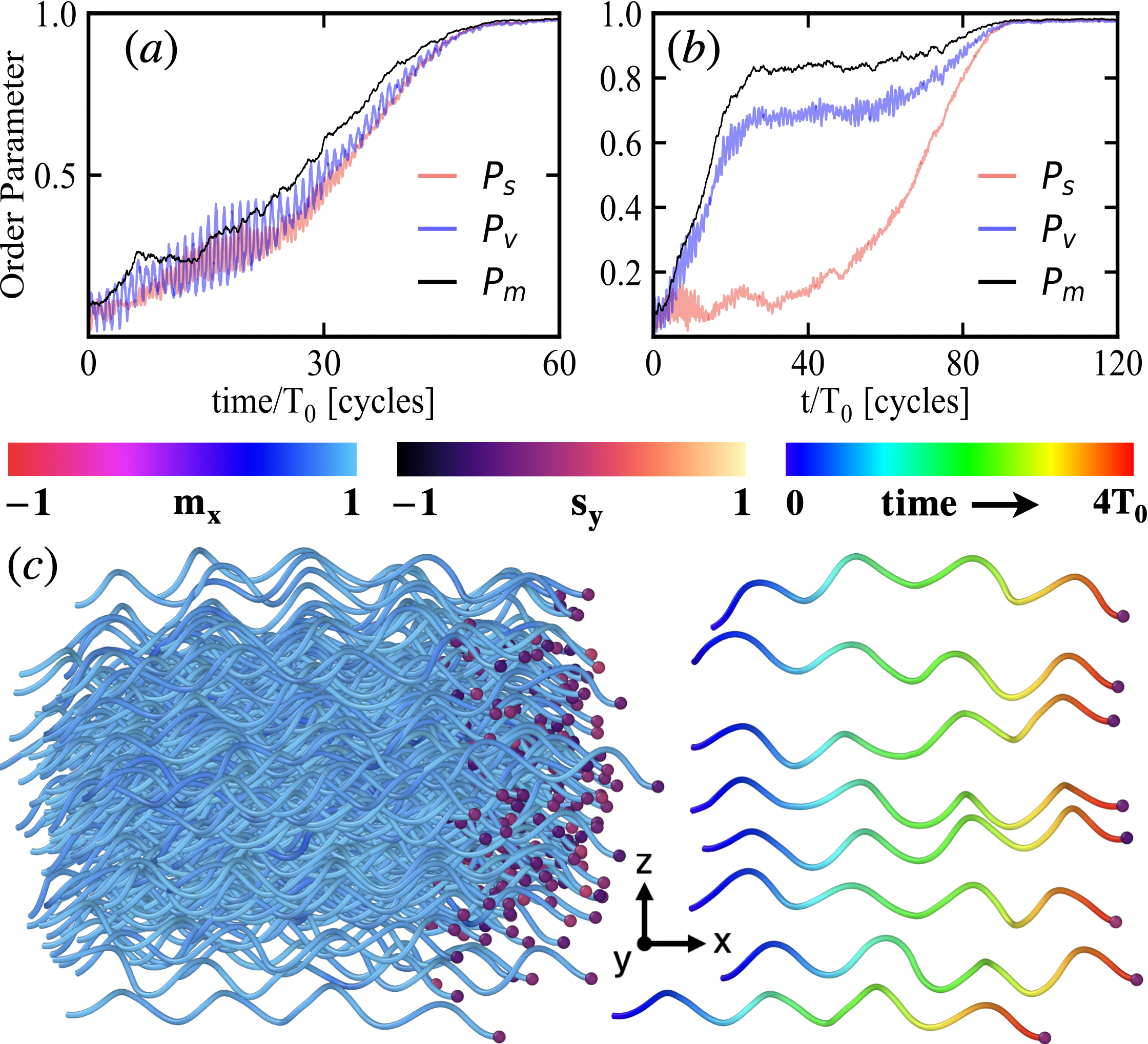}
\caption{Helical swimmers: Time-evolution of the order parameters $P_s$ (red), $P_v$ (blue) and $P_m$ (black) for (a)  $\phi\approx 0.083$, $\lambda \approx 0.19$ and (b) $\phi\approx0.125$, $\lambda \approx 0.1$. (c) Snapshots of the system in the synchronised state. The uwrapped trajectories of all the $N=286$ helical swimmers couloured according to $\bm{m}_x$ (left).  8 selected microswimmers with their trajectories coloured as a function of time (right). The particles are coloured according to $\bm{s}_y$. 
(The snapshots in (c) correspond to $\phi \approx 0.15$, $\lambda \approx 0.16$).}
\label{fig5}
\end{figure}

\paragraph*{Helical swimmers.---}
The helical swimmer trajectories are characterised by the ratio between the radius of curvature of the trajectory and the pitch length $\lambda = r_t/p$ (Fig.~\ref{fig1}b). The particle motion is 3-dimensional, leading to an increase of the probability of near-field interactions.
Hence, synchronisation is observed at lower $\phi$ than in the case of pure rotors (Fig.~\ref{fig2} and~\ref{fig4}). Similarly to circular swimmers, a high degree of order is observed in the synchronised state (Fig.~\ref{fig5}a and b), and the particles swim along a common direction, with their helical trajectories aligned (Fig.~\ref{fig5}c).
Interestingly, when the ratio $\widetilde{B}_{11}/B_1$ is decreased, the ordering dynamics is observed to change from a smooth growth to a two-step process where the velocity alignment  initially corresponds only to alignment in the polar direction (see {\it e.g.} Fig.~\ref{fig5}a and b, for $\lambda\approx 0.19$ and $\lambda \approx 0.1$, respectively).
Both the rotational frequency and the phase locking show comparable behaviour to the circular swimmers ~\cite{SupMat}.

When $\lambda =0$, the swimmers correspond to achiral neutral squirmers and the formation of pure polar order $(P_m>0;~P_s\sim 0)$ is observed  (blue diamonds in Fig~\ref{fig4}) in agreement with~\cite{evans2011orientational, alarcon2013spontaneous, yoshinaga2018hydrodynamic, yoshinaga2017hydrodynamic, delmotte2015large, theers2018clustering, oyama2017hydrodynamically}. Remarkably, we also find cases with $\lambda > 0$ with stable polar order  in the absence of azimuthal ordering ($P_m > 0;~P_s\approx 0$) (yellow diamonds in Fig.~\ref{fig4}). 

The synchronisation spans to low chiralities, and is observed for $\lambda \approx 0.03\ldots 0.22$ and $\phi \approx 2.5\ldots 20$\% (Fig.~\ref{fig4}). The $\lambda$ range corresponds to experimentally observed trajectories of biological swimmers such as $\lambda = r_t/p\approx 0.05$ for {\it T. thermophila}~\cite{marumo2021three} and $\lambda\approx 0.15$ for the 3-dimensional swimming of sperm~\cite{corkidi2008tracking}. We note that the transition between synchronised chiral states and the linear polar state ($\lambda=0$) is predicted to occur between $\lambda\lesssim 0.03$ and $\lambda=0$ (Fig.~\ref{fig4}). This suggests that synchronisation may well be observable at lower chiralities than $\lambda\approx 0.03$ considered in Fig.~\ref{fig4}.

\begin{figure}[t!] 
\includegraphics[width=0.48\textwidth]{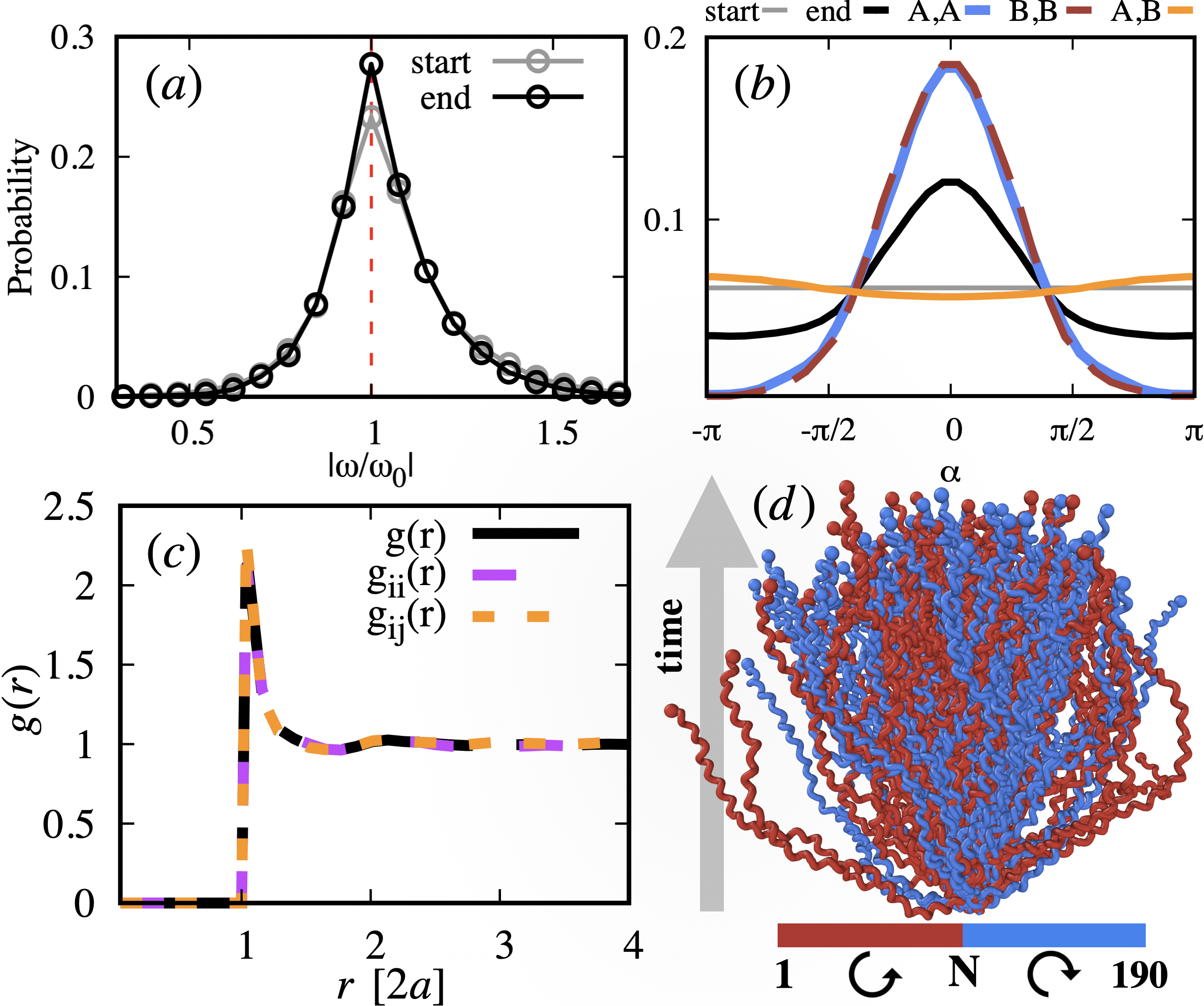}
\caption{Racemic mixture: Distributions of the (a) spinning frequency $\omega$ and (b) the phase-angle difference $\alpha$ calculated for all swimmer pairs (black), for clockwise (blue) and counter-clockwise (red) rotating populations, as well as for the cross population (orange). (c) Total (black), homochiral (violet) and heterochiral (orange) radial distribution functions.
(d) Snapshot of the unwrapped trajectories at the steady state after 20$T_0$. (The data corresponds to $\phi\approx 0.1$\ and $\lambda \approx 0.16$).}
\label{fig6}
\end{figure}

\paragraph*{Racemic mixture.---}
Finally, to study the effect of frustration, we construct a racemic mixture composed of right-handed and left-handed helical swimmers by choosing $C_1=\pm 0.001$ (Fig.~\ref{fig6}).
We start from a fully mixed isotropic state. At the steady state, the particles, on average, swim along a common direction (Fig.~\ref{fig6}d) and the rotational frequency $\omega$ is observed to peak at $\omega_0$ (Fig.~\ref{fig6}a).
The intra-species $\alpha$ shows strong phase-locking (blue and red curves in Fig.~\ref{fig6}b), whereas for the cross-species no significant peak is observed (orange curve Fig.~\ref{fig6}b), due to the oppositely spinning populations. However, the distribution shows a slight preference for $\alpha=\pm\pi$, which corresponds to a parallel orientation of the in-plane projections of the source dipoles~\cite{SupMat}.
Within the time-scale of the simulations, we observe no spatial separation of the swimmers --- the fluid-like pair-correlation function calculated within species and cross-species matches with the $g(r)$ of the whole system (Fig.~\ref{fig6}c).

\paragraph*{Conclusions.---}
Using hydrodynamic simulations we have investigated suspensions of microswimmers with chiral trajectories at the limit of zero thermal noise. The results suggest the emergence of hydrodynamic synchronisation as a naturally occuring collective phenomenon for microswimmers. The predictions should be relevant to a wide variety of experimental systems;  such as helically swimming bacteria~\cite{marumo2021three} and sperm~\cite{corkidi2008tracking}, or chiral Quincke rollers~\cite{zhang2020reconfigurable} and spherical ciliates~\cite{drescher2009dancing}, where rotational motion occurs naturally.
The observation of the intra-species synchronisation in the racemic mixture, provides a surprising example of two synchronised, interpenetrating, fluids. Further, it demonstrates that the synchronisation is maintained in the presence of hydrodynamic fluctuations arising from the source-dipole $1/r^3$ far-fields. This suggests, that it could be interesting to (re)analyse 3-dimensional correlations in the rotational degrees of freedom in systems exhibiting polar order, such as areas of uniform order in bacterial systems~\cite{dombrowski2004self, sokolov2007concentration, wolgemuth2008collective, cisneros2010fluid, zhang2010collective, sokolov2012physical, wensink2012meso, dunkel2013fluid, peng2021imaging} or polar flocks in motile colloids~\cite{bricard2013emergence}.

\begin{acknowledgments}
\paragraph*{Acknowledgments.} Discussions with Zaiyi Shen and Alois W\"urger are gratefully acknowledged. SS acknowledges University of Bordeaux, and the A. G. Leventis Foundation for funding, as well as cluster CURTA at MCIA for computational time. JSL acknowledges the French National Research
Agency (ANR) through Contract No. ANR-19-CE06-0012-01 and la region Nouvelle Aquitaine project GASPP for funding.
\end{acknowledgments}


\bibliography{paper}

\clearpage

\begin{appendix}

\clearpage



\onecolumngrid

{\center{\large {\bf  Supplementary material for Hydrodynamic synchronisation of chiral microswimmers} }} \\ 
\medskip
\begin{center}
    Sotiris Samatas and Juho Lintuvuori
\end{center}

\begin{center}
 Univ. Bordeaux, CNRS, LOMA, UMR 5798, F-33400 Talence, France 
\end{center}

\setcounter{equation}{0}
\setcounter{figure}{0}
\renewcommand{\thefigure}{S\arabic{figure}}
\renewcommand{\theequation}{S\arabic{equation}}
\medskip



\section{Additional details for the surface slip-flow}

The {\it purely tangential} slip velocity profile on the surface of a squirmer of radius $a$, found in an incompressible fluid at low Reynolds numbers, is given in spherical coordinates by an infinite series of modes for the polar and azimuthal components $\mathbf{e}_\theta$ and $\mathbf{e}_\varphi$\cite{pak2014generalized}:

\noindent
\begin{equation} \label{eq1}
     u_r\rvert _{r=a} = 0
\end{equation}
\begin{equation} \label{eq2}
\begin{gathered}
     u_\theta\rvert _{r=a} =\\ \sum_{n=1}^{\infty} \sum_{m=0}^{n} \Bigg[  
     \frac{-2\sin\!\theta P_n^{m{'}}}{na^{n+2}} ( B_{mn}\cos\!m\varphi + \widetilde{B}_{mn}\sin\!m\varphi )\\
     + \frac{m P_n^m}{a^{n+1}\sin\!\theta} (\widetilde{C}_{mn}\cos\!m\varphi - C_{mn}\sin\!m\varphi) \Bigg]
\end{gathered}
\end{equation}

\begin{equation}  \label{eq3}
\begin{gathered}
     u_\varphi\rvert _{r=a} =\\ \sum_{n=1}^{\infty} \sum_{m=0}^{n} \Bigg[  
     \frac{\sin\!\theta P_n^{m{'}}}{a^{n+1}} ( C_{mn}\cos\!m\varphi + \widetilde{C}_{mn}\sin\!m\varphi )\\
     + \frac{2m P_n^m}{na^{n+2}\sin\!\theta} (\widetilde{B}_{mn}\cos\!m\varphi - B_{mn}\sin\!m\varphi) \Bigg]
\end{gathered}
\end{equation}
\noindent

\noindent where $P_n^m = P_n^m(x)$ with $x=\cos\!\theta$ are the associated Legendre polynomials (with $n \geq 1$ and $0 \leq m \leq n$) and each mode can be identified by its corresponding coefficient: $B_{mn}$, $\widetilde{B}_{mn}$, $C_{mn}$ and $\widetilde{C}_{mn}$. The first two terms in the polar direction $(m = 0;~n=1,~2)$ correspond to the widely known squirmer model $u_\theta\rvert _{r=a}=B_1\sin\!\theta+\frac{1}{2}B_2\sin\! 2\theta$. 

\subsection{Choosing the parameters for chiral swimmers}

The modes with $n=1$ and $m=1$ are equivalent to the $n=1$ and $m=0$ modes discussed above but act in different directions, that is, their axis of symmetry is not the polar axis (or $z$-axis). The $B_{11}$ and $C_{11}$ modes are axisymmetric about the $x$-axis, and the $\widetilde{B}_{11}$ and $\widetilde{C}_{11}$ modes are axisymmetric about the $y$-axis. Therefore, the ``simplest" squirmer that does {\it not} swim in a straight line, consisting of just a hydrodynamic source dipole in the overall flow field, can be constructed by combining $B_1$ with $C_{11}$ or $\widetilde{C}_{11}$; $B_{11}$ with $C_1$ or $\widetilde{C}_{11}$; or $\widetilde{B}_{11}$ with $C_1$ or $C_{11}$. (Notice that using $B_1+C_1$, $B_{11}+C_{11}$, or $\widetilde{B}_{11}+\widetilde{C}_{11}$ would produce squirmers that swim in a straight line while spinning around their axis of symmetry). Each one of the six pairs of modes mentioned above leads to the squirmer having a circular trajectory in a given plane. Going one step further, helical motion can be produced by adding another $B$ mode in the direction perpendicular to the plane of the circular motion. What this effectively does is lead to a situation where the induced rotational and tranlational velocities of the squirmer are no longer perpendicular, $U\times\Omega\neq0$ and $U\cdot\Omega\neq0$, since the superposition of the two perpendicular source dipoles leads to a source dipole along the diagonal with an angle given by the relative strength of the initial dipoles.

Without loosing generality, in this work, we use $C_1+\widetilde{B}_{11}$ for circular swimmers; and $B_1+C_1+\widetilde{B}_{11}$ for helical swimmers (see Fig. 1 in the main text). Hence, the boundary conditions defining the slip velocity profile of our squirmer model are given in the polar $\hat{\bf{e}}_\theta$ and azimuthal $\hat{\bf{e}}_\varphi$ directions, by:

\noindent
\begin{equation} \label{eq4}
\begin{gathered}
     u_\theta\rvert _{r=a} = B_1\sin\!\theta + \widetilde{B}_{11}\cos\!\theta\sin\!\varphi\\
     u_\varphi\rvert _{r=a} = C_1\sin\!\theta + \widetilde{B}_{11}\cos\!\varphi , \ \ \ \ \ \ 
\end{gathered}
\end{equation}
\noindent

with $B_1$, $C_1$ and $+\widetilde{B}_{11}$ coefficients being our model parameters.

The flow field of the swimmers corresponds to a neutral squirmer (source dipole) $\mathbf{B}$ rotating around the polar axis $\bm{m}$ at an angle $\psi = |\tan^{-1}(\widetilde{B}_{11}/B_1)|$ with angular velocity $\omega_0 = C_1/a$. The magnitude $B=\sqrt{\widetilde{B}_{11}^{2} + B_1^2}$, gives a single isolated particle swimming speed $v_0=\tfrac{2}{3}B$.  $\psi = 90^\circ$ corresponds to circular swimmers and $\psi = 0^\circ$ to linear squirmers. The radius of {\color{black} curvature} of the trajectory $r_t$, is  given by $r_t=2\widetilde{B}_{11}a/(3C_1)$  and the period by $T_0=2\pi/\omega_0=2\pi a/C_1 = 2\pi r_t/\widetilde{B}_{11}$. The helical pitch corresponds to the distance $p$ travelled along the direction given by the particle's polar axis during a rotational period $T_0= 2\pi a/C_1$ and is given by $p = 4\pi B_1 a/(3C_1)$. To characterise the helical swimmers $0^\circ<\psi<90^\circ$, we define a ratio between the radius of curvature and the pitch length $\lambda = \frac{r_t}{p}=\frac{\tilde{B}_{11}}{2\pi B_1}$, which gives: $\psi =|\tan^{-1}2\pi\lambda|$.


\section{Simulation details}
The microswimmers are modelled as spherical squirmers, radius $a$, using lattice Boltzmann method~\cite{Juho_SM_2016,Zaiyi_EPJE_2018}.
The no-slip boundary condition at the particle surface~\cite{ladd1994numerical1, ladd1994numerical2, nguyen2002lubrication}, is modified to take into account the active slip-flows~\cite{llopis2010hydrodynamic, pagonabarraga2013structure}. 

We use lattice units where distance is given by the lattice spacing $\Delta x=1$ and time in simulation time-steps $\Delta t = 1$. Particles of radius $a=8$ (diameter $\sigma=16$) are placed within a cubic lattice $L_x=L_y=L_z=160$ with periodic boundary conditions, corresponding to a volume $V=160^3$. A short-range repulsive interaction is implemented to avoid particle overlaps~\cite{Juho_SM_2016,Zaiyi_EPJE_2018} with a cut-off distance $1\Delta x$.

The fluid density is set to $\rho=1$ and dynamic viscosity to $\mu=0.5$. We set $C_1=0.001$ leading to an intrinsic angular velocity of $\omega_0 = C_1/a =1.25 \cdot 10^{-4}$ and to a rotational Reynolds number $Re_\Omega = \rho \omega_0 a^2/\mu = 0.016$. To realise different trajectory radii the source dipole strength is varied $\widetilde{B}_{11} \in [0,0.007]$. These correspond to swimming speeds of $v_0 = 2/3 \widetilde{B}_{11} \in [0,4.67 \cdot 10^{-3} ]$, (linear) Reynolds numbers $\mathrm{Re} = \rho v_0 a/\mu \in [0,0.075]$ and $r_t = v_0 a/C_1 \in [0,4.67a]$. 

For the helical swimmers we use $B_1=0.005$ {\color{black} and vary $\widetilde{B}_{11} \in [0,0.007]$ corresponding to a maximum $\mathrm{Re}\approx 0.09$. The parameters give a pitch length $p=4\pi B_1 a/3C_1\approx 21a$ for the data in Figures 4-6 in the main text.}

For the initial configurations, the squirmers are homogeneouly and isotropically distributed in the simulation box. The simulations are run for a minimum time $t_{max}=500T_0$ (where $T_0 = 2\pi/\omega_0$ is the intrinsic period of a single rotor) corresponding to $\sim25\times10^6\Delta t$ LB steps. 
For dilute suspensions (packing fractions up to $10\%$), we use $t_{max}=1000 T_0$ ($\sim50\times10^6\Delta t$ LB steps).

\section{Mapping to SI units}
Assuming a particle radius $1\mu$m and using the viscosity of water $10^{-3}$Pas as well as typical $\mathrm{Re}\sim 10^{-2}$, a single lattice length $\Delta x$ and time $\Delta t$ can be mapped to $\sim 0.1\mu$m and $\sim 10\mu$s, respectively. Therefore a typical simulation run of $\sim 10\times 10^{6}$ LB steps 
corresponds to $100$s in real time. 

\clearpage

\section{Additional details for calculating the random close packing of the trajectories for circular swimmers}

\begin{figure}[h!] 
\includegraphics[width=0.98\textwidth]{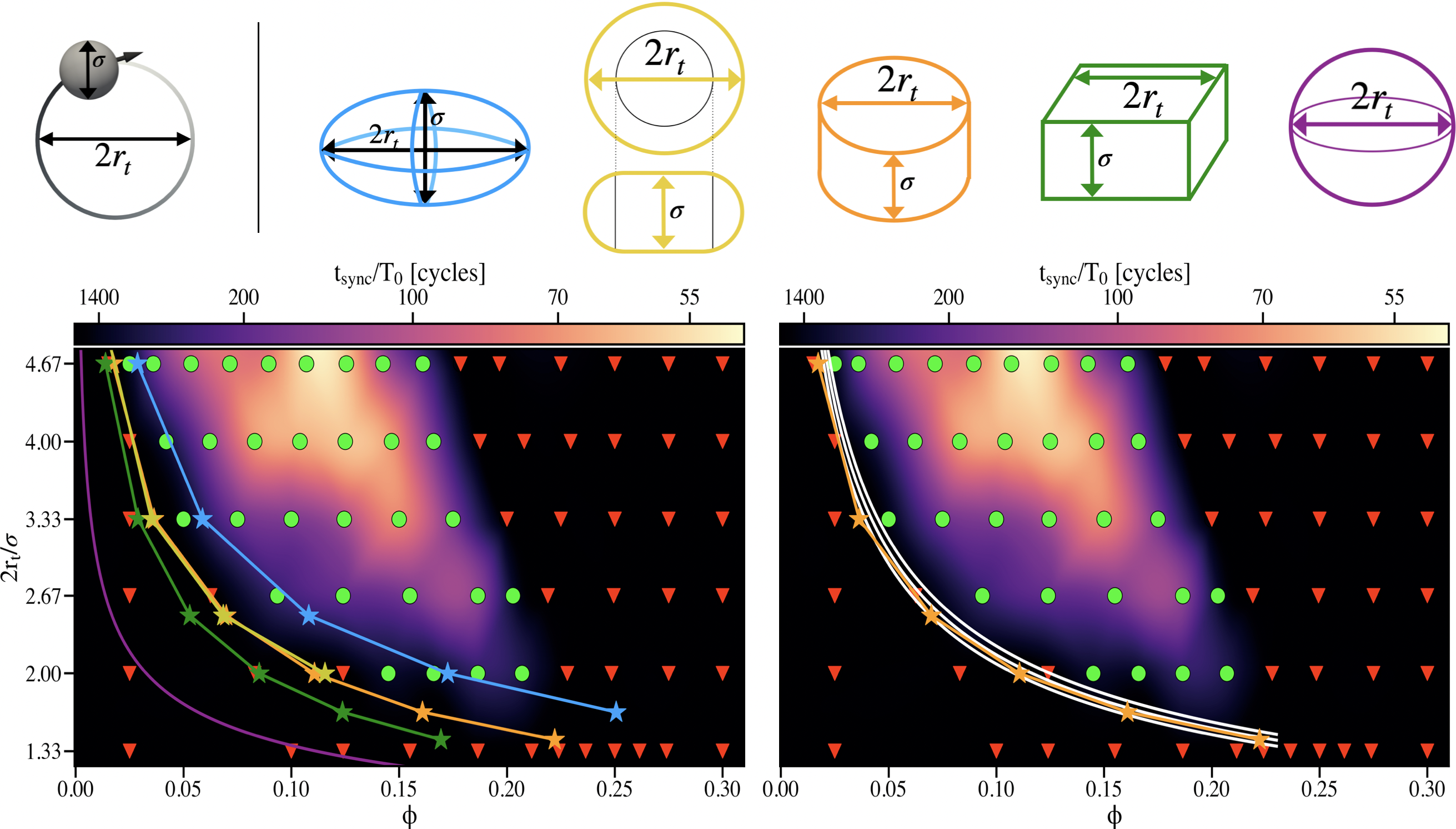}
 
\caption{The steady state diagram of the circular swimmers. Bottom left panel: mapped fits of different shapes (top panel) at their random close packing $\phi_c'$ (according to eq.\ref{eqS1}). Oblate ellipsoids corresponding to the average values of ref~\cite{donev2004improving, zhou2013discrete} (blue), oblate hard spherocylinders --- OHSC ref.~\cite{martinez2009simulation} (yellow), discotic cylinders ref.~\cite{li2010maximum, liu2018evolutions} (orange), rectangular cuboids ref.~\cite{liu2017maximally} (green) and spheres ref.~\cite{scott1969density, berryman1983random, torquato2000random} (purple). All oblate shapes, with volumes $\mathrm{v}' \sim r_t^2 \sigma $ give a reasonable fit to the data, while spheres, with $\mathrm{v}' \sim r_t^3$, do not. Bottom right panel: random close packing for discotic cylinders in orange just as before, according to ref.~\cite{li2010maximum, liu2018evolutions}, and assuming a constant $\phi_c'$ (white) for: $\phi_c' = 75$\%, $70$\% and $65$\%. We use $\phi_c' = 70\%$ in the main text.} 
\label{figSIV}
\end{figure}

{\color{black}
The formation of polar order, with linear neutral squirmers has been attributed to aligning near-field hydrodynamic interactions~\cite{yoshinaga2017hydrodynamic,yoshinaga2018hydrodynamic}.
In accordance with~\cite{yoshinaga2017hydrodynamic,yoshinaga2018hydrodynamic}, we argue that near-field interactions are important for the the hydrodynamic synchronisation as well. Considering the circular swimmers with a trajectory radius $r_t$, if the volume fraction $\phi$ is below a threshold value, the circular swimmers are far enough from each other so that the near-field hydrodynamic interactions are negligible, and there will be no aligning interaction between them to eventually lead to the synchronised state. For $t>T_0$, an isolated circular swimmer with a diameter $\sigma$ encircles an area $A = \pi r_t^2$ (in the plane perpendicular to the polar axis). Therefore, when ${r_t {\sim} {\sigma}}$, a circular swimmer can be thought to occupy an effective volume: $\mathrm{v}_{\mathrm{eff}} \approx \pi r_t^2\sigma$, where $\sigma= 2 a$ is the diameter of the particle.

Consequently, we can expect the aligning interaction between two swimmers to arise when there is an overlap of their respective effective volumes $\mathrm{v}_{\mathrm{eff}}$. The validity of this argument can be tested by studying the random close packing $\phi_c'$ of different oblate geometrical objects with a volume $\mathrm{v}'$ and an aspect ratio $w = \sigma/(2r_t)$. Fig.~\ref{figSIV} shows the $\phi$ curves, given by equations~\ref{eqS1}, corresponding to the \textit{mapped} $\phi_c'$ for systems with the same number density $N/V$ and with particle volumes $\mathrm{v}'$: Oblate ellipsoids \cite{donev2004improving, zhou2013discrete} (blue), oblate hard spherocylinders --- OHSC \cite{martinez2009simulation} (yellow), discotic cylinders \cite{li2010maximum, liu2018evolutions} (orange), rectangular cuboids \cite{liu2017maximally} (green) and spheres using random close packing of $64$\%~\cite{scott1969density, berryman1983random, torquato2000random} (purple). The lower boundary data for the synchrosnisation, is generally well fitted with disk-like objects $w\sim\sigma/(2r_t)< 1$, while spheres $w=1$ (purple) fail to do so.

The mapping for each curve appearing in the steady state diagram is given by: 
\begin{equation} \label{eqS1}
     \phi = \phi_c' \frac{\mathrm{v}}{\mathrm{v}'}
\end{equation}
\noindent where $\mathrm{v}=4/3 \pi a^3$ is the actual volume a single spherical swimmer and $\phi_c'$ is the random close packing for the corresponding shape with a volume $\mathrm{v}'$ at the same number density. 
 
In the main text (Fig. 2) we  fit the boundary using discotic cylinders ($\mathrm{v}'=\pi r_t^2 \sigma$) at $\phi_c' = 70\%$. The bottom right panel in Fig.~\ref{figSIV} shows the data using the random close packing of discotic cylinders $\phi'_c(w)$ (orange) from references~\cite{li2010maximum, liu2018evolutions}. The three white curves use a constant $\phi_c' = 65\%$, $\phi_c' = 70\%$ and $\phi_c' = 75\%$ from left to right, respectively.

We note that the above reasoning is only valid for circular swimmers with a reasonably small radius of curvature $r_t\sim a$. When $r_t>>a$ the trajectories could easily interpenetrate without affecting the particle dynamics.
}

\clearpage

\section{Additional figures for circular swimmers: MSD and full snapshots}

\begin{figure}[h] 
\includegraphics[width=0.98\textwidth]{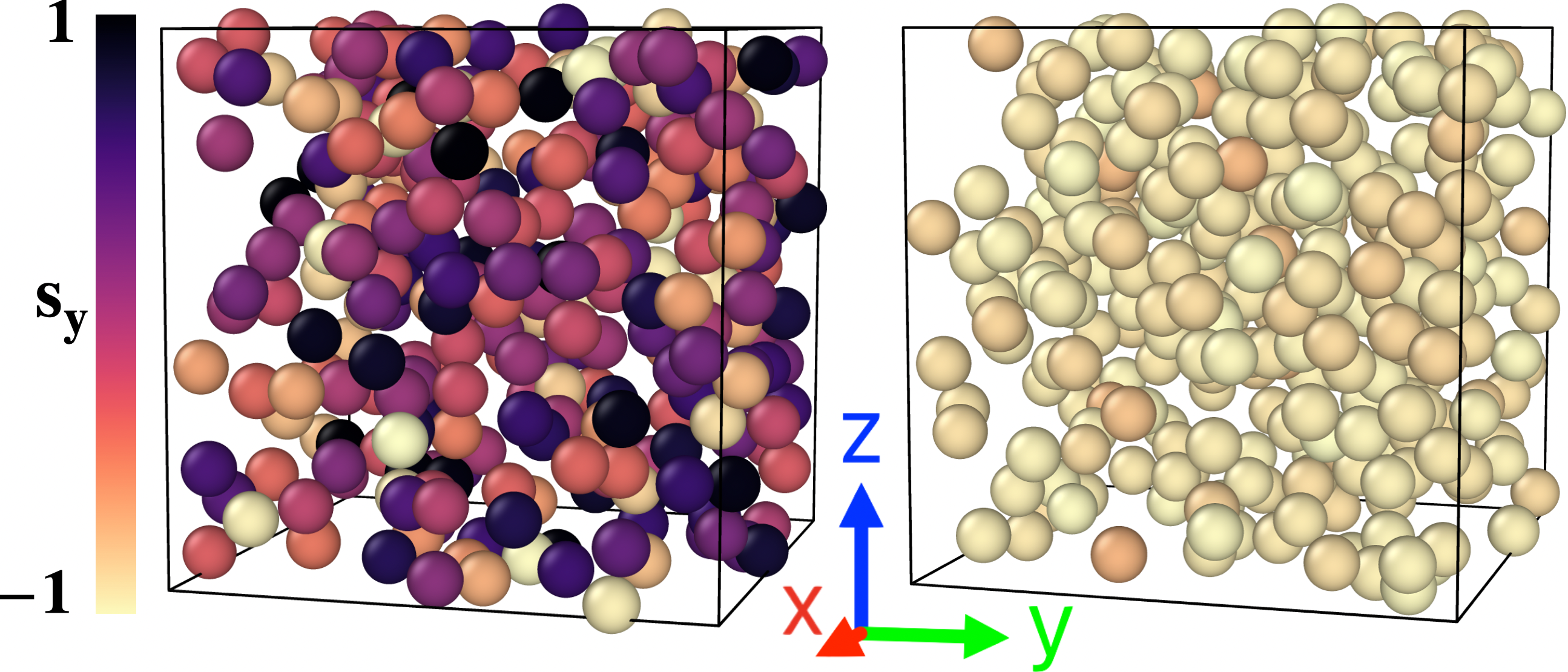}
\caption{Snapshots of the system ($\phi=0.15$ and $r_t=3.33a$ in the steady state diagram for circular swimmers) at the beginning (left) and end (right). The $N=286$ particles are coloured according to the orientation of their $\bm{s}$ vector along the {\it lab frame} $y$-axis. The full spectrum of colours is present in the system at the beginning indicating an isotropic state, whereas in the synchronised state (right), all the particles are yellow indicating the high degree of alignment present in the system (Fig. 3 in main text). } 
\label{figSV1_2}
\end{figure}

\begin{figure}[h] 
\includegraphics[width=0.98\textwidth]{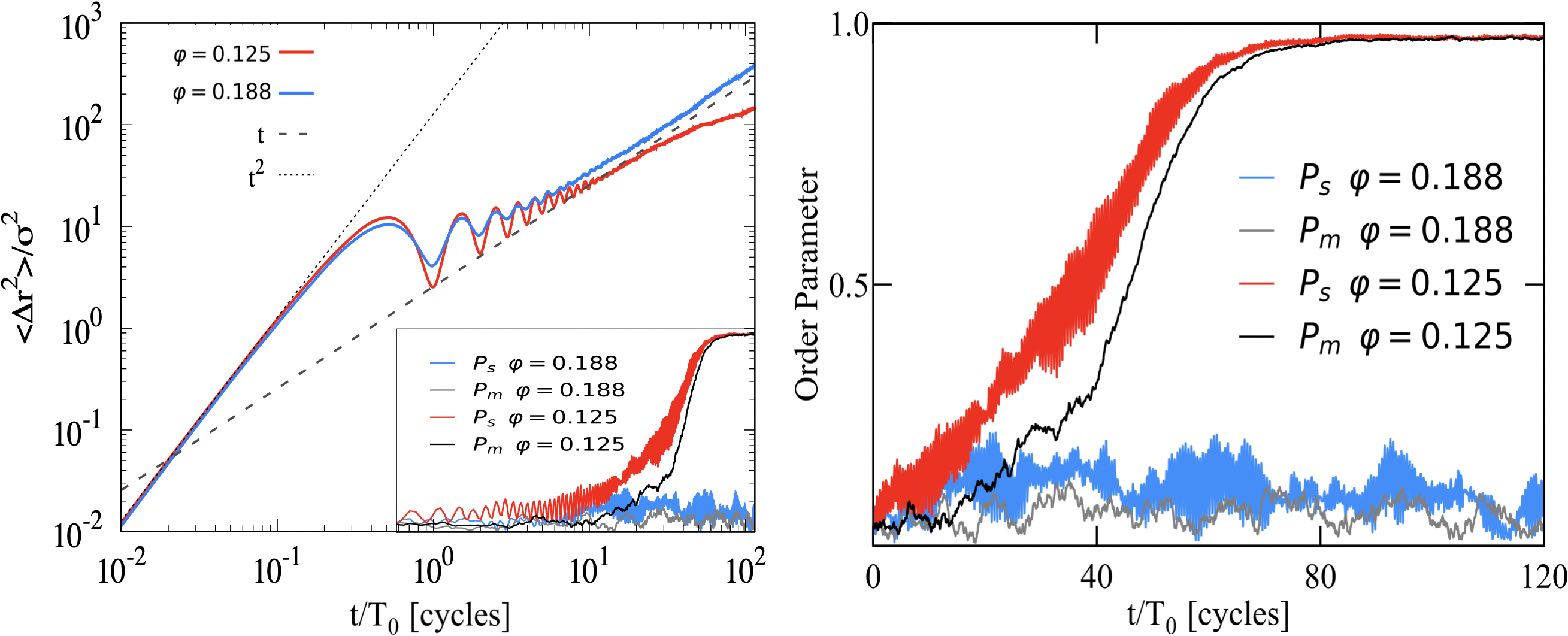}
\caption{Left panel: The mean square displacment (MSD) for the the high density isotropic state (blue) and synchronising state (red) for circular swimmers with $r_t\approx 4a$. At short times, the ballistic regime corresponds to the the swimmers completing one cycle during their intrinsic period $T_0$. At long times, the active isotropic state shows diffusion MSD$\sim t$ (blue), while at the synchronised state the dynamics becomes sub-diffusive MSD$\sim t^\alpha$ with $\alpha< 1$ (the red curve at long times). The inset shows the corresponding azimuthal $P_s$ and polar $P_m$ order parameters (in the same log-log scale). Right panel: $P_s(t)$ and $P_m(t)$ in linear scale.} 
\label{figSV2}
\end{figure}

\begin{figure}[h] 
\includegraphics[width=0.98\textwidth]{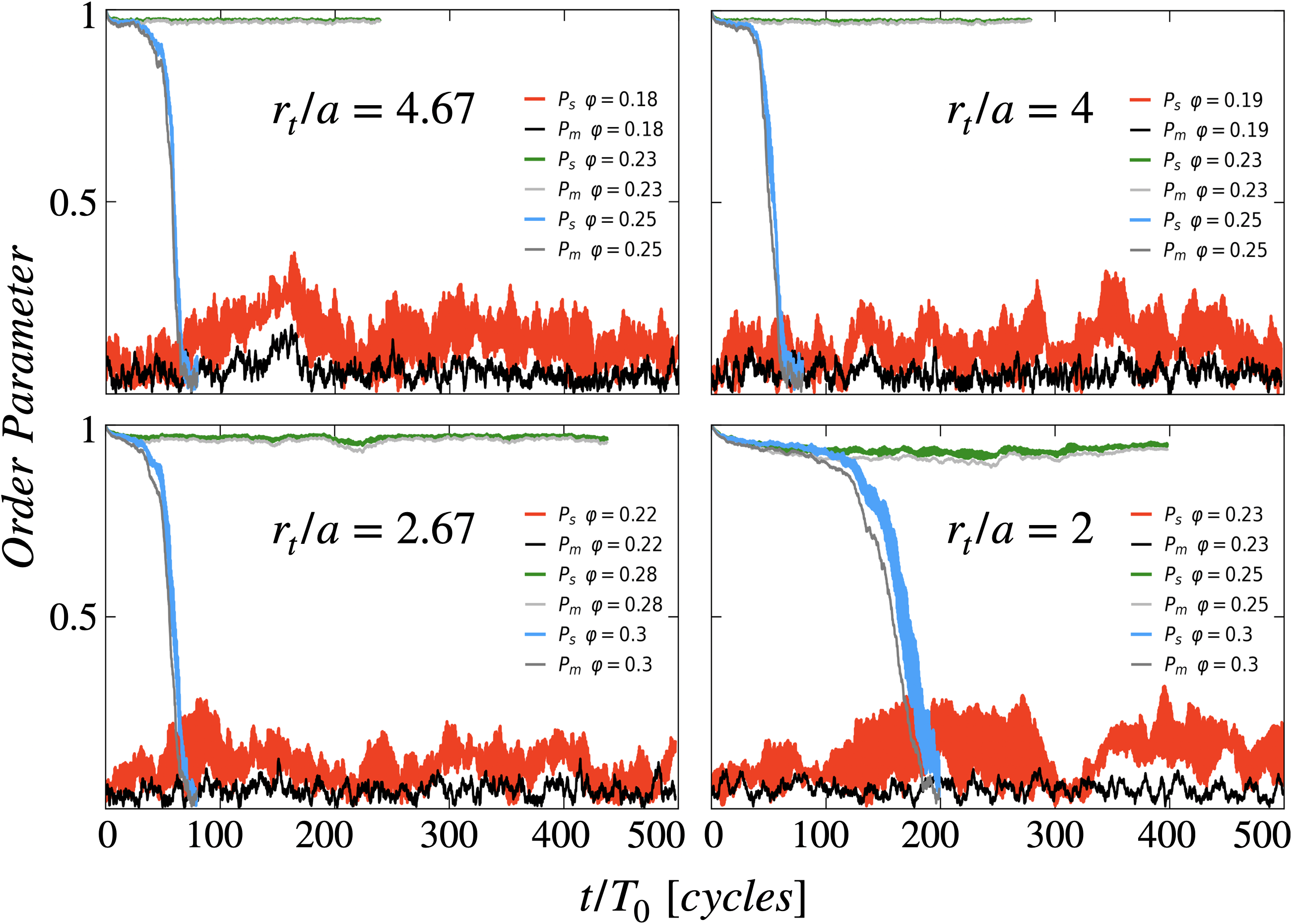}
\caption{Comparison between starting the simulations form an isotropic initial state (as used in the main text) and fully ordered states for different $\phi$ at constant $r_t$. The system shows hysteresis-type behaviour within the simulation time, near the high density transition line.} 
\label{figSV3}
\end{figure}

\pagebreak

\clearpage

\section{Additional figure for the helical swimmers: Distribution of the spinning frequencies and phase angles}

\begin{figure}[h!] 
\includegraphics[width=0.98\textwidth]{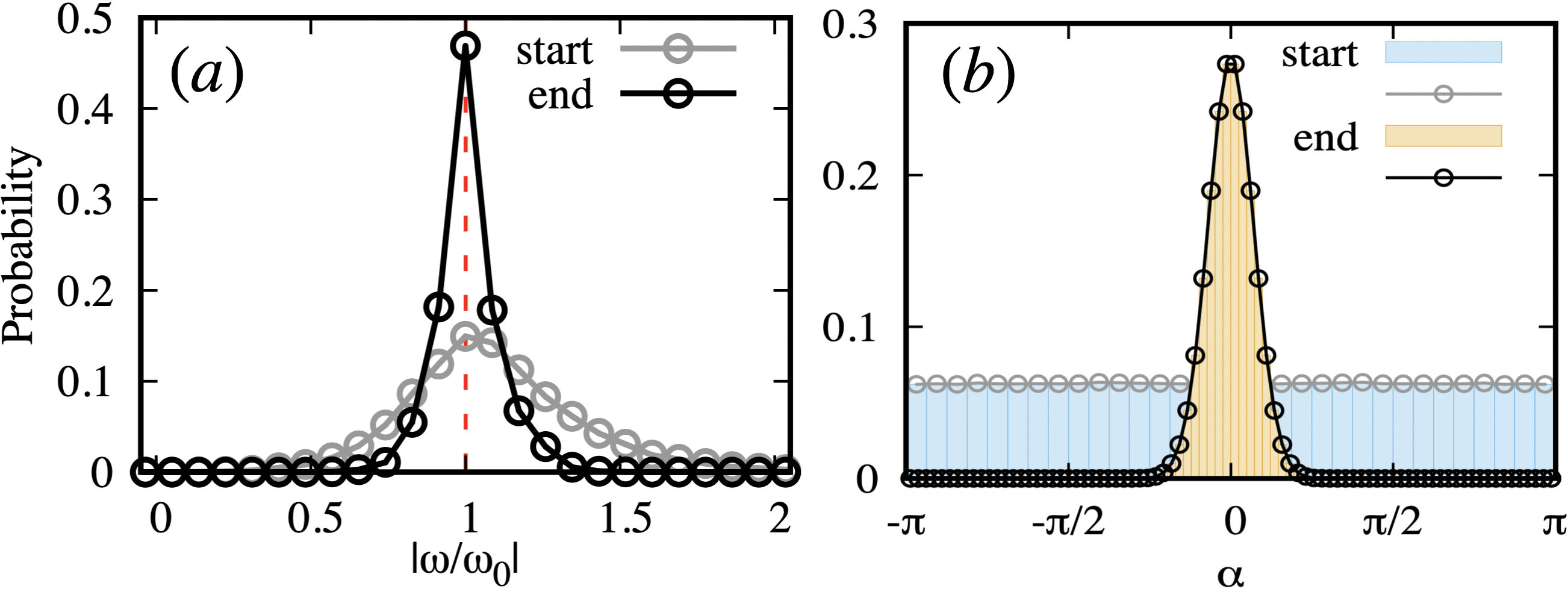}
\caption{Probability distribution for $N=286$ helical swimmers corresponding to $\phi \approx 0.15$, $r_t \approx 3.33a$, $\lambda \approx 0.16$ and $\psi \approx 45^{\circ} $ of the (a) angular velocities $\omega$ and (b) phase lag angle $\alpha$ between all particle pairs, at the start and end of the simulation.} 
\label{figSVI}
\end{figure}

\clearpage

\section{Additional details for the racemic mixture}

\begin{figure}[h!] 
\includegraphics[width=0.7\textwidth]{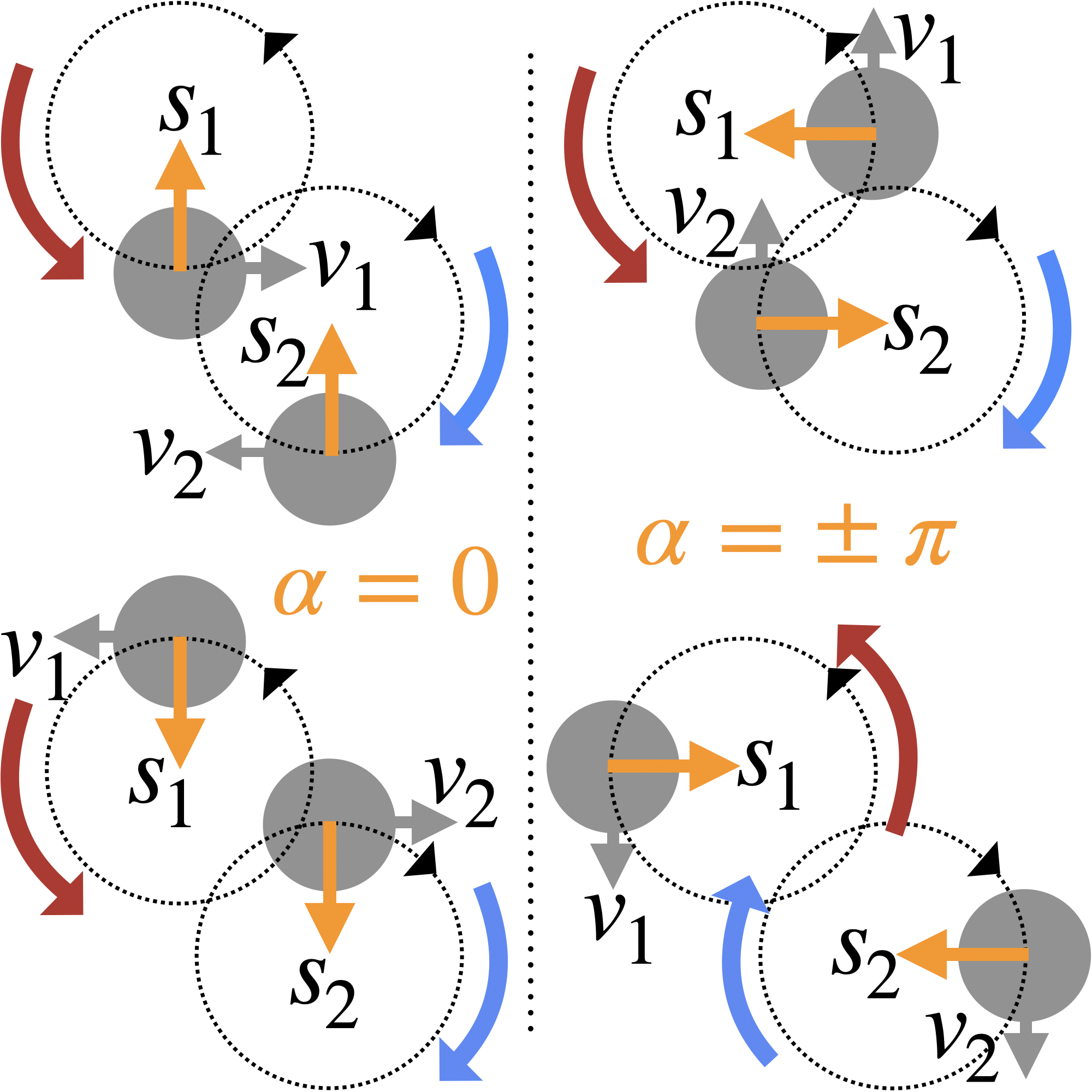}
\caption{Schematic showing the parallel ($\alpha=0$) and anti-parallel ($\alpha=\pm\pi$) configurations of a pair of rotors with opposite chirality in the plane perpendicular to the average global polar director $\mathbf{P}_M$. When the lag-angle $\alpha = 0$, the in-plane velocities $\mathbf{v}_1$ and $\mathbf{v}_2$ (given by the gray arrows) are antiparallel. For $\alpha=\pm \pi$ the velocites are aligned, corresponding to a polar state in the plane perpendicular to the rotational (polar) axis. The arrows $\mathbf{s}_1$ and $\mathbf{s}_2$ mark the direction of the assigned azimuthal directors. The polar directors $\mathbf{m}_1$ and $\mathbf{m}_2$ are perpendicular to the plane shown.} 
\label{figSVII}
\end{figure}

\clearpage


\end{appendix}
\end{document}